\documentclass[9pt,twocolumn,twoside]{pnas-new}

\templatetype{pnasresearcharticle} 
\hypersetup{
    colorlinks,
    linkcolor={red},
    citecolor={green},
    linkcolor={black}
}
\title{Surface Densities Prewet a Near-Critical Membrane}

\author[a,b]{Mason Rouches}
\author[c]{Sarah Veatch} 
\author[b,d]{Benjamin Machta}

\affil[a]{Department of Molecular Biophysics and Biochemistry, Yale University, New Haven, Connecticut 06511, USA}
\affil[b]{Systems Biology Institute, Yale University, West Haven, Connecticut 06516, USA}
\affil[c]{Department of Biophysics, University of Michigan, Ann Arbor, MI 48109, USA}
\affil[d]{Department of Physics, Yale University, New Haven, Connecticut 06511, USA}

\leadauthor{Rouches} 

\correspondingauthor{E-mail: mason.rouches@yale.edu, sveatch@umich.edu, benjamin.machta@yale.edu}

\begin{abstract}
Recent work has highlighted roles for thermodynamic phase behavior in diverse cellular processes. Proteins and nucleic acids can phase separate into three-dimensional liquid droplets in the cytoplasm and nucleus and the plasma membrane of animal cells appears tuned close to a two-dimensional liquid-liquid critical point. In some examples, cytoplasmic proteins aggregate at plasma membrane domains, forming structures such as the post-synaptic density and diverse signaling clusters. Here we examine the physics of these surface densities, employing minimal simulations of co-acervating polymers coupled to an Ising membrane surface in conjunction with a complementary Landau theory. We argue that these surface densities are a novel phase reminiscent of pre-wetting, in which a molecularly thin three-dimensional liquid forms on a usually solid surface. However, in surface densities the solid surface is replaced by a membrane with an independent propensity to phase separate. We show that proximity to criticality in the membrane dramatically increases the parameter regime in which a pre-wetting-like transition occurs, leading to a broad region where coexisting surface phases can form even when a bulk phase is unstable. Our simulations naturally exhibit three surface phase coexistence even though both the membrane and the polymer bulk can only display two phase coexistence on their own. 
We argue that the physics of these surface densities enables diverse functions seen in Eukaryotic cells.
\end{abstract}

\dates{\today}

\begin{document}

\maketitle
\thispagestyle{firststyle}
\ifthenelse{\boolean{shortarticle}}{\ifthenelse{\boolean{singlecolumn}}{\abscontentformatted}{\abscontent}}{}
Eukaryotic cells are heterogeneous at scales far larger than individual macromolecules, yet smaller than classically defined organelles. Proteins, RNA, and DNA can self-organize into three-dimensional, liquid-like droplets in the cytoplasm and nucleus~\cite{Alberti2017} and lipids and proteins in the plasma membrane similarly organize into two-dimensional domains, often termed `rafts'~\cite{Sengupta07}.
These domains and droplets are thought to form in part due to a thermodynamic tendency of their components to phase separate into coexisting liquids. 
Proteins and other molecules within three-dimensional droplets are held together through weak but specific multi-valent interactions~\cite{Brangwynne09,Li2012,Priftis2012}. 
Lipids and other membrane components interact through less specific effective forces that arise from hydrophobic mismatch, from the interaction of lipid headgroups and from steric interactions between lipid tails~\cite{Honerkamp-Smith09}. Cell derived vesicles separate into coexisting phases termed liquid-ordered ($l_o$) and liquid-disordered ($l_d$), passing through a critical point when cooled somewhat below growth temperature~\cite{Veatch08}. At growth temperatures, domains arising from proximity to this critical point likely resemble corresponding low temperature phases at small scales but with finite size and lifetime. 

Some \textit{surface densities} appear to form due to a combination of these forces. In these systems proteins aggregate in a thin film at a membrane surface with some components strongly attached to membrane lipids~\cite{Banjade2014,Su2016,Zeng2016,Beutel2019,Wu2019} while others are free to exchange with the bulk. The protein components of these films can phase separate in the bulk, but only at substantially higher concentrations than are seen in vivo~\cite{Zeng2018,Banjade2014,Beutel2019}. 
Examples of these surface densities include the Nephrin/Nck/NWasp system that plays a role in cell adhesion~\cite{Banjade2014,Case2019}, T-cell signaling clusters~\cite{Su2016}, and the post-synaptic density~\cite{Zeng2016,Zeng2018}. 


Systems that phase separate in three-dimensions can undergo wetting transitions~\cite{deGennes1985,Bonn2009}, 
where there is a change in the bulk phase that adheres to a surface. 
In addition to wetting transitions which take place inside of coexistence on the bulk phase diagram, surfaces of bulk fluids can undergo \textit{prewetting} transitions~\cite{Cahn1977,Nakanishi1982} which occur near to bulk coexistence. In prewetting transitions, a normally unstable bulk phase is stabilized through favorable interactions with a surface, leading to a surface film which resembles the nearby (in the thermodynamic phase diagram) bulk phase, but which is molecularly thin. 

The behavior of membrane domains and protein droplets have both been successfully described using theories of phase transitions in fluid systems~\cite{Brangwynne2015,Honerkamp-Smith09}, but there has been less work interpreting these surface aggregates. 
We use lattice Monte-Carlo simulations in conjunction with a minimal Landau theory to explore the physical principles governing these droplets. We argue that surface densities are similar to prewet phases, but with subtlety arising from their adhesion to a two-dimensional liquid which is itself prone to phase separating. We predict a novel surface phase sensitive to both membrane and bulk parameters which we argue describes a wide variety of structures which are already biochemically characterized.


\section*{Simulation results}
\noindent \textbf{Model overview-}
In our simulations we describe the membrane using a conserved order parameter two-dimensional square-lattice Ising model~\cite{Honerkamp-Smith08,Machta11}. In this model spins roughly represent membrane components - proteins or lipids - which prefer the liquid-ordered (spin up, white in figures) or liquid-disordered (spin down, dark) membrane phases. The Ising model introduces two parameters, the coupling between neighboring spins $J_{mem}$, and $M$, the difference in the number of up and down spins. Experiments suggest that plasma membrane composition is tuned close to the critical point of de-mixing which occurs in the Ising model when $M=0$ (equal number of up and down spins) at a critical coupling $J=J_c$.     

We model phase-separating cytoplasmic proteins as a lattice coacervate. Here two types of polymers, each 20 monomers in length, live on a 3D cubic lattice. Unlike polymers interact attractively with coupling $J_{bulk}$ (in the range of $ k_{B}T$, see methods for exact values of all parameters) when occupying the same lattice position, and like polymers cannot occupy the same position. We also include a weak non-specific nearest-neighbor interaction between all polymer sites which allows phases to localize in space~\cite{Freeman_Rosenzweig2017,Xu2020}. The two polymer types roughly represent interacting components of phase-separated droplets in the cytoplasm or oppositely charged, synthetic polymers such as poly-lysine and poly-glutamine~\cite{priftis2012-1,priftis2012-2}. In synthetic systems, the coupling between polymers can be modulated by salt and polymer length. Cells alter their coupling through post-translational modifications, changes in salt, pH, and changes in valency ~\cite{Snead2019,Shin2017}. 

To couple our membrane and bulk models we introduce tethers which may be thought of as membrane-localized proteins. Tethers connect to up-spins on the membrane and extend several (five) lattice spacings into the third dimension where they interact with bulk polymers through an attractive interaction $J_{tether}$. Unlike bulk polymers, tethers translate in just two-dimensions across the membrane surface. In cells tethers correspond to lipidated or transmembrane proteins that interact with proteins in the cytoplasm~\cite{Jiang2018}. In synthetic systems tethers have been engineered through strong non-covalent binding attachment of peptides or proteins to lipid headgroups~\cite{Banjade2014,Su2016,Zeng2018}

\begin{figure}[!htb]
    \centering
    \includegraphics[width=\columnwidth]{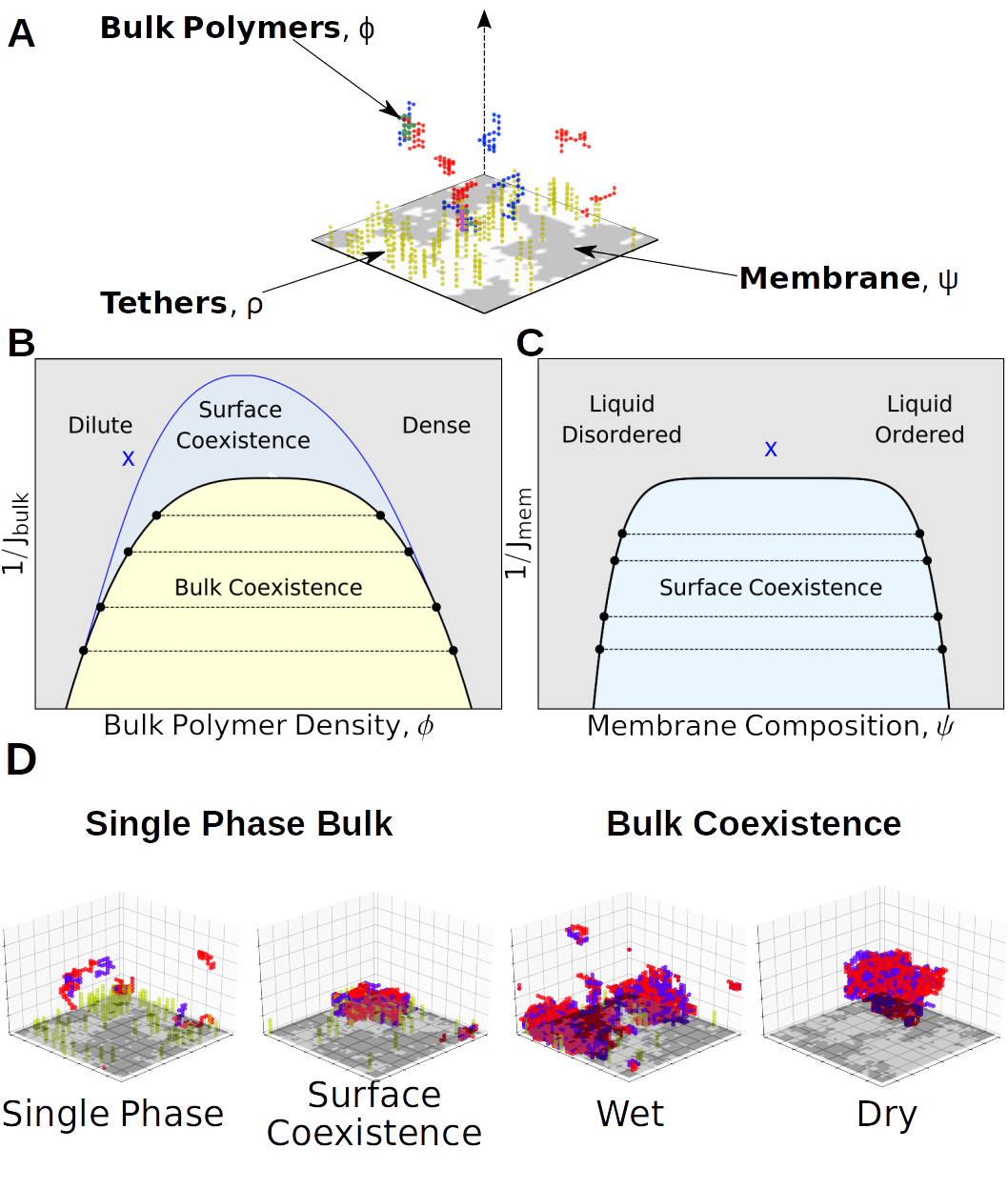}
    \caption{\textbf{Bulk and Surface Phases}: \textbf{A}) Cartoon of the minimal model used to describe surface densities. In our simulations red and blue lattice polymers have an attractive interaction in the three-dimensional bulk. An Ising model on the bulk's boundary contains bright/dark pixels representing liquid ordered/disordered preferring components of a membrane. Tethers (yellow polymers) are connected to up spins, and have an attractive interaction with components of the bulk. \textbf{B}) Schematic Bulk Phase diagram. On a plot of inverse interaction strength (like temperature) vs polymer density, the bulk phase diagram contains a single bulk phase region (blue and grey) and a region where a dense and dilute phase coexist (yellow). The shape of the bulk coexistence curve does not depend on location within the surface phase diagram. \textbf{C}) The region of surface coexistence depends on both bulk and surface parameters. In B and C, we show two two-dimensional slices where surface coexistence  occurs in the blue region. The blue X in B corresponds to location in the bulk phase diagram for which the surface phase diagram is plotted in C; moving this location would change the shape of the blue coexistence curve in C. The blue X in C is the location in the surface parameters for which the surface phase diagram is plotted in B. \textbf{D}) Example phases observed in Monte-Carlo simulations. On the left are two examples without bulk coexistence, one with surface coexistence, one without. On the right are two examples of coexisting bulk phases, a wet phase adheres to the membrane and a dry phase avoids the membrane. }
    \label{fig:SchematicCartoon}
\end{figure}

\noindent \textbf{Bulk phase behavior is independent of surface details-}  We expect the 3D bulk polymers to have a phase diagram which, in the thermodynamic limit, does not depend on properties of the Ising surface. In the absence of a  membrane, at fixed polymer number, the bulk can can be in either a uniform state or can display coexistence between a polymer dense state and polymer dilute state. A phase diagram for this is sketched in figure \autoref{fig:SchematicCartoon}B in black; at low coupling, $J_{bulk}$, or high temperature, the state is uniform for any bulk concentration. At higher coupling there is a coexistence region where tie-line endpoints, the black circles in \autoref{fig:SchematicCartoon}B, represent physically accessible polymer densities and where both endpoints have the same chemical potentials and Gibbs Free energies. To observe coexistence we perform simulations at fixed polymer number with equal numbers of red and blue polymers. While the coexistence region of the composition-coupling plane does not depend on the properties of the membrane surface, it's appearance in simulation does;  in a `dry' regime, the polymer dense droplet avoids the surface, while in a `wet' regime it adheres to at least a portion of the surface. Wetting transitions occur when the bulk phase which adheres to the surface changes - here this can be achieved by altering either the bulk properties or the membrane properties, and in particular by changing the concentration of tethers. Our focus, however, is on the surface phases which can coexist even in a single phase region of the bulk. Henceforth we conduct simulations in which bulk polymers are instead held at fixed chemical potential in the dilute regime (see methods for parameter values).

\noindent \textbf{Multiple surface phases can coexist on the boundary of a single bulk phase-}
In the absence of tethers, the membrane can phase separate if the interaction strength $J_{mem}$ is lower than a critical value. In this sense, it is possible for the system to display surface phase coexistence even when the bulk is uniform. In the absence of tethers our membrane's phase diagram is well characterized, with a large coexistence region. When tethers are added which prefer one of these two phases, we qualitatively see that the bulk polymer distribution is different near these two phases (see~\autoref{fig:SchematicCartoon}D). This implies that bulk properties should be able to qualitatively change the surface phase diagram even in the absence of bulk phase separation. In particular, increasing bulk coupling should be able to induce phase separation at the surface even when membrane interactions are too weak to induce phase separation on their own ($J_{mem}<J_{c,mem}$, equivalent to $T > T_{c,mem}$). 

We thus expect the surface phase diagram to depend on parameters of the bulk polymer solution and on the membrane and tethers which make up the surface. We sketch two 2-dimensional slices through this five-dimensional phase diagram in \autoref{fig:SchematicCartoon}B,C. At a given point in the bulk phase diagram (blue x in \autoref{fig:SchematicCartoon}B) we see a surface coexistence region resembling that for a two-dimensional coexisting liquid prone to phase separating via an Ising transition (blue shaded region in \autoref{fig:SchematicCartoon}C). Alternatively, by fixing the surface parameters at the blue x in \autoref{fig:SchematicCartoon}C, the surface coexistence region is plotted in \autoref{fig:SchematicCartoon}B.

These surface coexistence regions are analogous to prewetting where, for example, a liquid film adheres to a solid surface of a gas phase bulk. In these classical examples there can be either an abrupt or a continuous transition to a prewet state triggered by either increasing bulk density, or by lowering temperature. In the limit where $J_{mem}=0$ our system is analagous to this, albeit with the additional complexity of a fluid surface quantity in membrane tethers. More substantially different, the Ising model on the surface also participates in the prewetting transition by further enhancing the interactions that drive surface aggregation.

\begin{figure}[htb!]
    \centering
    \includegraphics[width=\columnwidth]{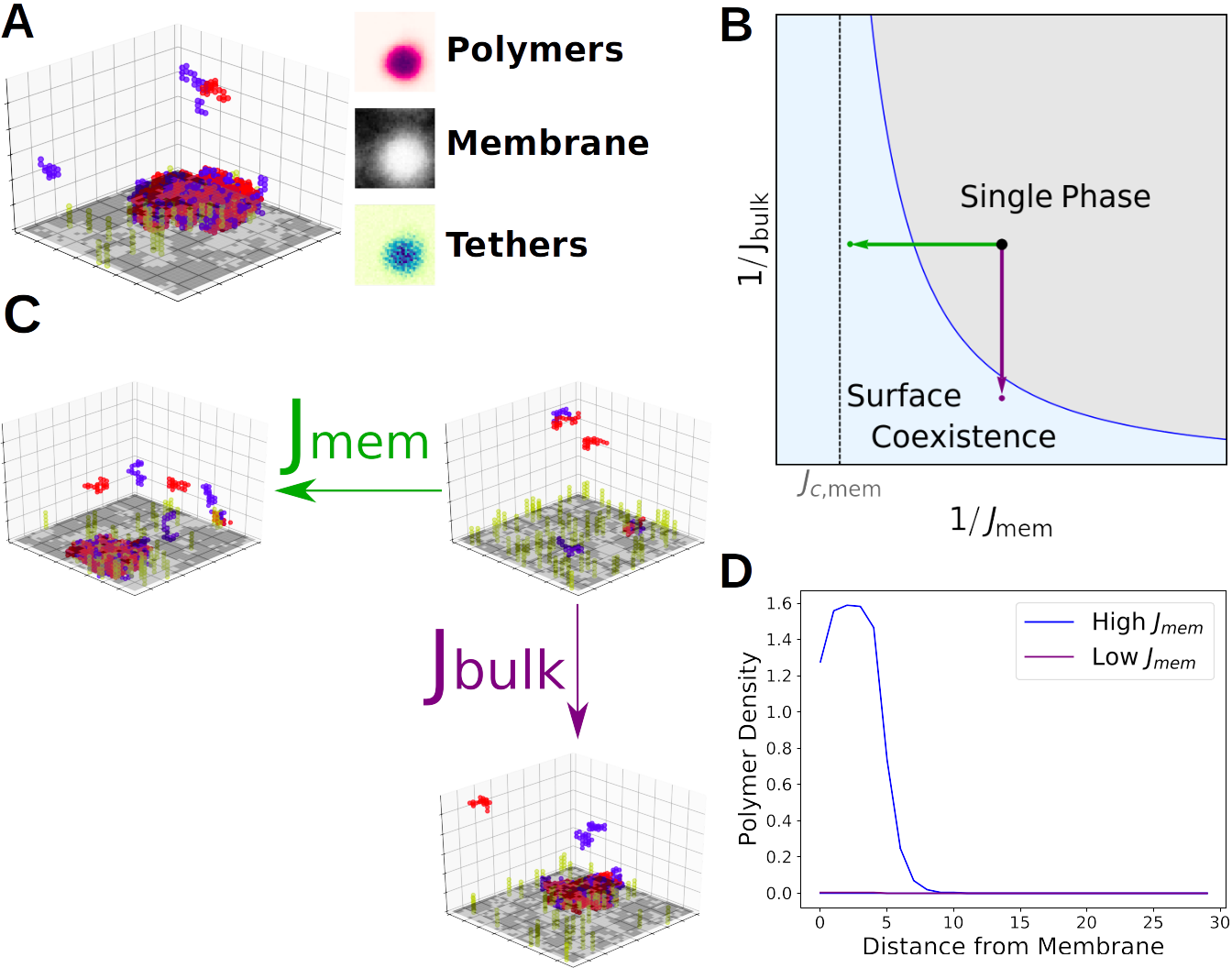}
    \caption{\textbf{Prewetting of surface densities:} \textbf{A}) Snapshot of a simulation where a polymer-dense droplet prewets the membrane surface even though droplets are unstable in the bulk. Time averaged membrane, tether, and polymer compositions are shown at right. \textbf{B}) Schematic phase diagram in terms of membrane and bulk couplings. 
    A single phase system (black dot) can move into the surface coexistence region by increasing $J_{bulk}$ (purple arrow) or increasing $J_{mem}$ (green arrow). \textbf{C}) Simulations at weak bulk and membrane coupling are brought into the coexistence region through increasing $J_{bulk}$ (purple) or $J_{mem}$ (green). \textbf{D}) Density of polymers as a function of distance from the membrane. A system at weak bulk and membrane couplings sees a single phase (purple, simulation in lower left of C) while systems at stronger membrane couplings see two coexisting polymer density profiles (blue, simulation in right of C).}
    \label{fig:SchematicPhaseDiagram}
\end{figure}

\noindent \textbf{Surface and bulk properties together determine the surface phase diagram-}
To more quantitatively explore the surface phase diagram in simulation we found a region of parameter space that displays two coexisting phases far from their critical point so that phases could be easily identified in small simulations (see \autoref{fig:SchematicPhaseDiagram}A). These two phases differ from each other in their membrane order, their density of tethers and the density of polymers near them. We expect to be able to move from a single phase to two-phase coexistence by increasing either membrane interactions or bulk interactions (schematically shown in \autoref{fig:SchematicPhaseDiagram}B). This is demonstrated in \autoref{fig:SchematicPhaseDiagram}C; a single phase surface is brought into the surface coexistence region by increasing the coupling between bulk polymers (lower) or by increasing the interactions between membrane components (left). Each of these coexisting surface phases has a characteristic polymer density profile with distance from the surface (\autoref{fig:SchematicPhaseDiagram}D). 
Although we primarily focus on membrane and bulk couplings, we confirmed that prewetting can additionally be triggered by increasing the number of tethers on the membrane (see Supplement). 


Our simulation results suggests that the range of $J_{bulk}$ in which we see prewetting expands significantly as the membrane critical temperature is approached (see \autoref{fig:SchematicPhaseDiagram}B) or as we bring the membrane towards it's critical composition ($M=0$) at fixed coupling strength (see Supplement). These results imply that the membrane critical point expands the surface coexistence region, which we explore more quantitatively using a Landau theory below. 

\begin{figure}[htb!]
    \centering
    \includegraphics[width=\columnwidth]{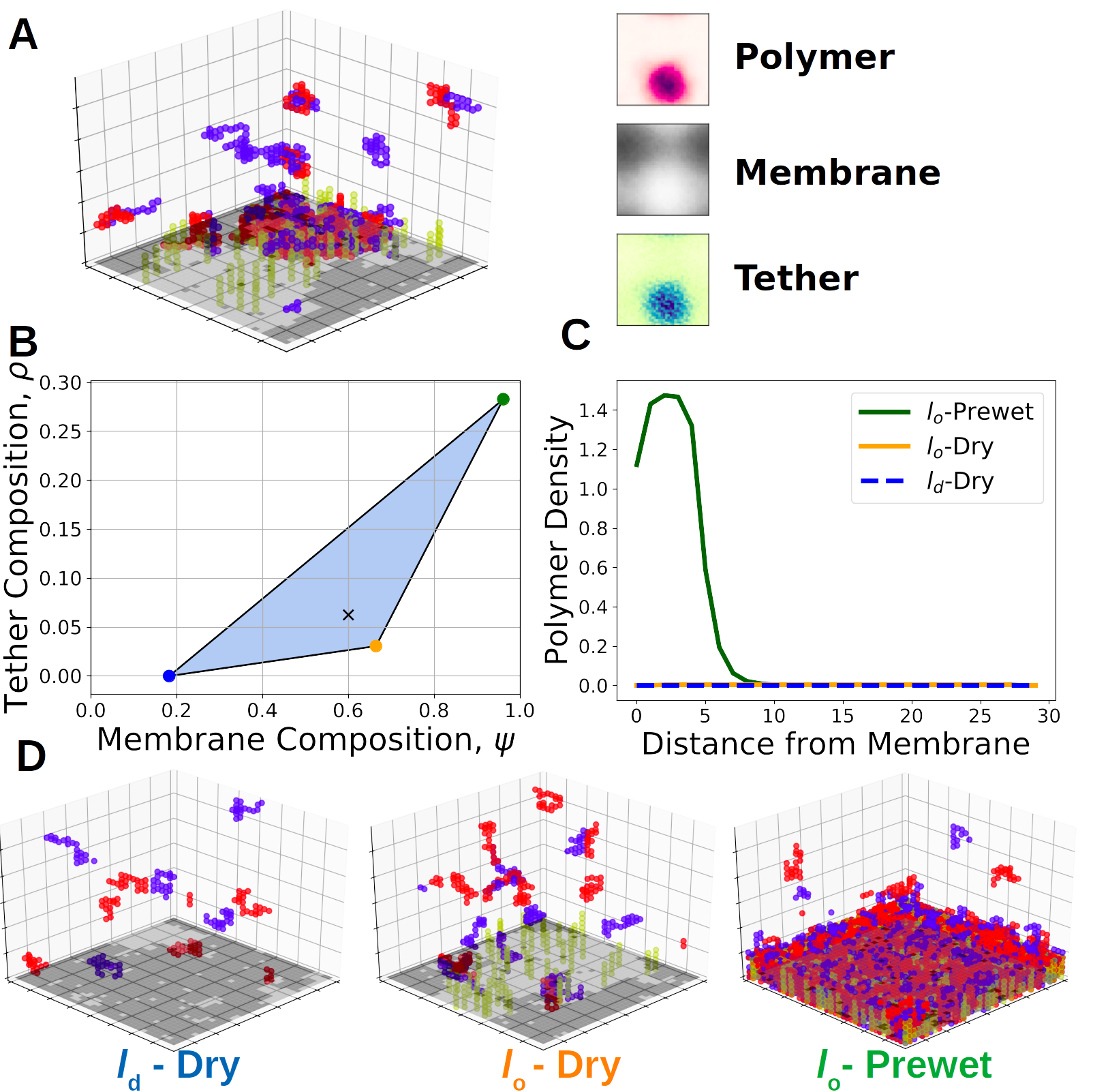}
    \caption{\textbf{Three Phase Coexistence:} \textbf{A}) Simulations display three-phase coexistence where a polymer-dense droplet prewets a phase-separated membrane. Views of time-averaged tether density, membrane composition, and polymer density show a tether and polymer-dense phase rich in ordered components, an ordered membrane phase with a small amount of tethers, and a disordered membrane phase devoid of tethers. \textbf{B}) Phase diagram over membrane and tether composition extracted from the simulation in A. Membrane and tether compositions falling inside the blue triangle split into three-phases, each with a composition given by the vertices of the triangle. Black X corresponds to the surface composition of the simulation in A. \textbf{C}) Polymer density profiles, as a function of distance from the membrane in each of the three phases. \textbf{D}) Snapshots of simulations ran at compositions corresponding to the endpoints of three-phase coexistence.}
    \label{fig:3PhaseCoex}
\end{figure}

\noindent \textbf{Simulations demonstrate three-phase surface coexistence-}
Three phase coexistence in our model is allowed by Gibbs phase rule; two conserved quantities on the surface - tether and membrane composition - allow for up to $2+1$ phase coexistence. Indeed, we see three coexisting surface phases in simulations (\autoref{fig:3PhaseCoex}A) each with distinct membrane compositions, as well as tether and polymer density profiles. Three phase coexistence generally occurs at polymer couplings that would prewet a single-phase membrane and at membrane couplings that would phase separate even in the absence of any bulk coupling. We extracted the tether and membrane composition of each phase, plotted in \autoref{fig:3PhaseCoex}B. When tether and membrane composition lay inside the shaded triangle the system phase separates into phases with tether and membrane compositions given as the vertices of the triangle, each with an accompanying density profile shown in \autoref{fig:3PhaseCoex}C. We ran simulations at each of these surface compositions to observe individual phases, shown in \autoref{fig:3PhaseCoex}D.

We describe surface phases by their membrane and polymer compoitions. What we denote the $l_o$-Prewet phase is composed of a $l_o$-like membrane rich in tethers, with an adhered polymer droplet. The $l_d$-Dry phase is an $l_d$ membrane excluded of tethers and with lacking an enhancement of bulk polymers. The final phase, $l_o$-Dry, consists of an $l_o$-like membrane somewhat sparse in tethers and without a significant enrichment of bulk components. Here we assume tethers prefer $l_o$ lipids;  $l_d$ preferring tethers would instead form an analogous $l_d$-Prewet phase.


\section*{Landau analysis of Surface Phase behavior}
 
 Our lattice simulations serve to give a primarily qualitative and intuitive picture for the phases we see. To more quantitatively understand these surface phases we  introduce a Landau free-energy functional, modifying the analysis commonly used to theoretically describe prewetting transitions to incorporate membrane and tethers. As in standard analysis we introduce order parameter fields, and a Landau functional of their configuration, and consider the order parameter of the system to take the configuration which globally minimizes the Landau functional~\cite{Goldenfeld1992}. Phase coexistence occurs when two configurations of fields both have the same minimum value of the free energy. 
 
Our Landau functional, $\mathcal{L}$, describes a bulk system ($z>0$) with a surface at $z=0$, with $\vec{x}$ parameterizing the plane parallel to the surface. A single bulk order parameter $\phi(\vec{x},z)$ describes the local density of polymers while two surface order parameters, $\rho(\vec{x})$ and $\psi(\vec{x})$ describe the density of tethers and the membrane composition along an $l_o$-$l_d$ tie-line. We define $\phi_0(\vec{x})=\phi(\vec{x},z=0)$ and, suppressing coordinates, we write a Landau free energy for this system as $\mathcal{L} = \mathcal{L}_{3D} + \mathcal{L}_{2D}$ with:
\begin{align}
\mathcal{L}_{3D} &= \int_{V} d^2\vec{x} dz\text{ } \frac{1}{2}(\nabla\phi)^2 + f_{3D}\left(\phi \right) \\
\mathcal{L}_{2D} &= \int_{\partial V} d^2\vec{x} \text{ } f_{2D}\left(\psi, \rho, \phi_0 \right) \nonumber
\end{align}
Where $f_{2D}$ and $f_{3D}$ describe the energy of the surface and bulk systems: 
\begin{align}
    f_{3D}(\phi) &= \frac{t_{bulk}}{2}\phi^2 + \frac{u_{bulk}}{4!}\phi^4 - \mu_{bulk}\phi \nonumber\\
    f_{2D}(\psi,\rho) &= \underbrace{\frac{t_{mem}}{2}\psi^2 + \frac{u_{mem}}{4!}\psi^4 - \lambda_{\psi}\psi}_{f_{membrane}}   \\& + \underbrace{\frac{(\rho - \rho_{\star})^2}{2\rho_{\star}} + \frac{(\rho - \rho_{\star})^4}{12\rho_{\star}^3} - \lambda_{\rho}\rho}_{f_{tether}} - \underbrace{h_{\psi}\rho\psi - h_{\phi}\rho\phi_0}_{f_{int}} \nonumber
\end{align}
Where $t_{bulk}$ is the distance from the bulk critical point, $\mu_{bulk}$ the chemical potential of the bulk system, and $t_{mem}$ the distance from the membrane critical point. $u_{bulk}$ and $u_{mem}$ are higher order membrane and bulk couplings. The first two terms in $f_{tether}$ are taken from an expansion of $\rho\log{\rho}$ at $\rho_{\star}$. Lagrange multipliers $\lambda_{\psi}$ and $\lambda_{\rho}$ enforce membrane and tether composition, respectively. Membrane-tether and tether-bulk interactions are set by $h_{\psi}$ and $h_{\phi}$. We take $h_{\phi} > 0$ and $\mu_{bulk} < 0$, corresponding to a dilute-phase polymer mixture whose components interact favorably with tethers. 
Minimizing this Landau functional determines the value of two derivatives and a functional derivative, $\partial \mathcal{L}/\partial\psi = \partial \mathcal{L}/\partial\rho =0$ and $\delta \mathcal{L}/\delta\phi(z)=0$.

In the thermodynamic limit the 2D Ising model and tethers act as a boundary condition for the bulk, and thus cannot influence which bulk phases are stable. The bulk phase is the value of $\phi_\infty$ which globally minimizes $f_{3D}(\phi)$, defining $f_{bulk}=f_{3D}(\phi_\infty)$. The resulting bulk phase diagram recapitulates~\autoref{fig:SchematicCartoon}A, but with mean field exponents. At high temperatures, or low concentration of polymers there is a single dilute phase, which can coexist with a second dense phase at lower temperature. 

\begin{figure}[htb!]
    \centering
    \includegraphics[width=\columnwidth]{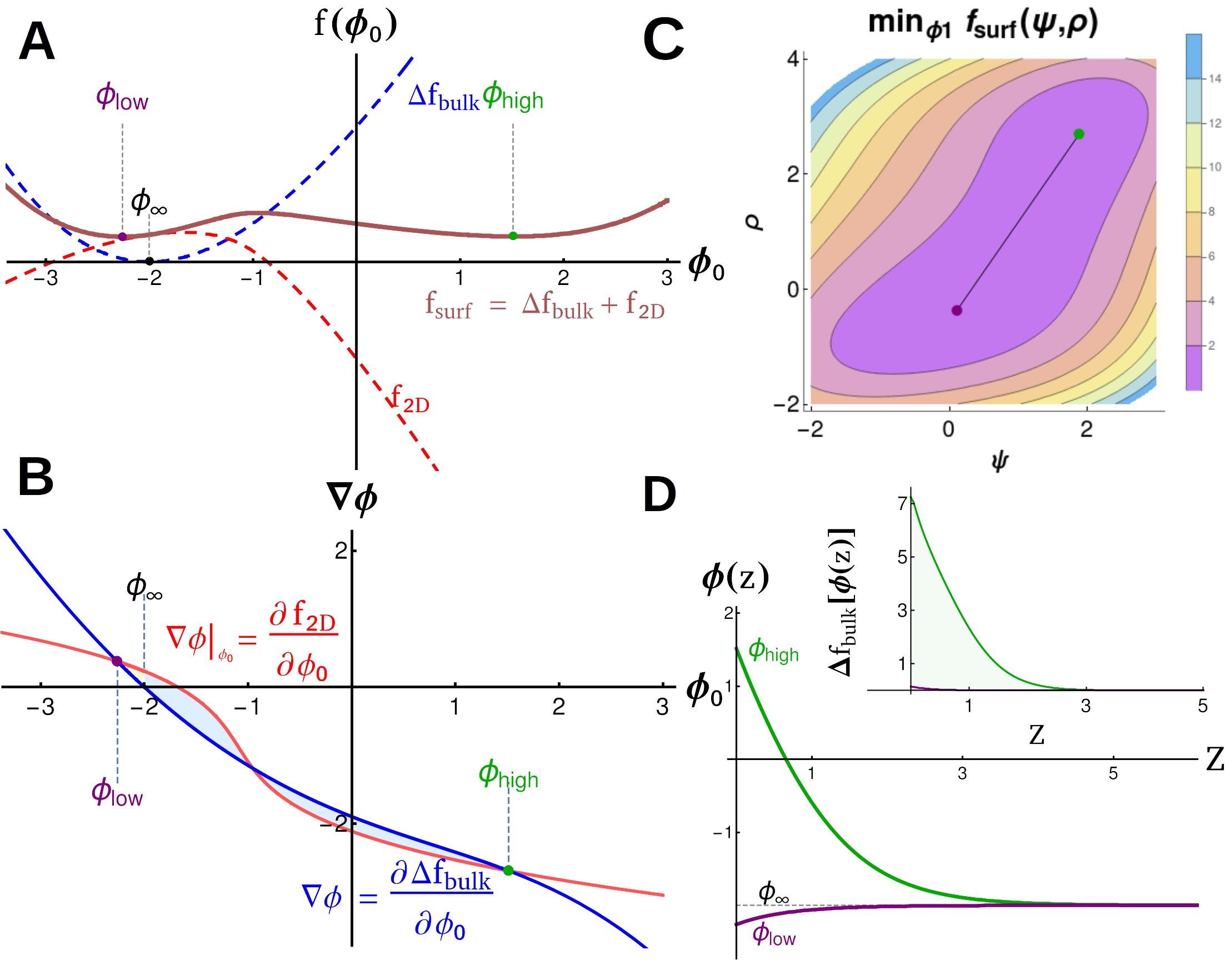}
    \caption{\textbf{Landau Theory of surface phases:} \textbf{A}) Bulk $\Delta f_{bulk}$ (blue), membrane $f_{2D}$ (red, already minimized over $\psi, \rho$), and surface $f_{surf} = \Delta f_{bulk} + f_{2D}$ (brown) free energies as a function of surface polymer density $\phi_0$. There are two energy minimia, $\phi_{low}$ and $\phi_{high}$ in the combined $f_{surf}$ even in the absence of multiple minima in $\Delta f_{bulk}$ or $f_{2D}$. \textbf{B}) Gradient construction used to visualize solutions. Intersections of derivatives of $f_{2D}$ (red) and $\Delta f_{bulk}$ (blue) give possible surface solutions $\phi_{low}$ and $\phi_{high}$. The free energy difference between these solutions is given by the area between these curves, visualized as the shaded regions. Changing the position or slope of surface or bulk lines changes the surface solutions. \textbf{C}) Surface free energy $f_{surf}$ calculated over values of $\psi$ and $\rho$, minimized first over $\phi_0$. Two minima (purple and green) correspond to surface compositions that minimize the free energy of the membrane and tethers along with their resulting contributions to bulk energy. \textbf{D}) Density profiles and energy density (inset) as a function of distance from the membrane $z$ for the two physical phases. Both $\phi_{high}$ and $\phi_{low}$ decay to the bulk density $\phi_{\infty}$. This adds unfavorable contributions to the free energy $\Delta f_{bulk}(\phi(z))$ that are balanced by contributions from $f_{surf}$}
    \label{fig:LandauTheory}
\end{figure}

\noindent \textbf{Analysis of Surface behavior-}
Outside of bulk coexistence, $\mathcal{L}_{3D}$ is globally minimized by a unique $\phi(\vec{x},z)=\phi_\infty$, where $\mathcal{L}_{3D}=V f_{bulk}$, with $V$ the system volume and where $A$ is its area. The free energy of the surface, $f_{surf}$, contains membrane contributions and contributions from surface induced distortions of the bulk field $\Delta f_{bulk}$:
\begin{align}
    \mathcal{L}_{surf} &= \mathcal{L}- V f_{bulk}= A f_{surf}(\rho,\psi,\left\{ \phi(z)\right\}) \nonumber \\
f_{surf} &=f_{2D}(\rho,\psi,\phi_0) + \underbrace{\int dz \frac{1}{2}(\nabla\phi)^2 + f_{3D}\left(\phi \right) - f_{bulk}}_{\Delta f_{bulk}}
\end{align}
For a given location in the bulk phase diagram the surface can exhibit its own set of phases and transitions which are local minima of $f_{surf}$. While $\Delta f_{bulk}$ and  $f_{2D}$ cannot be independently minimized, they can be independently minimized for a given value of $\phi_0$. Local minima of $f_{2D}|_{\phi_0}$ satisfy the conditions that $\partial f_{2D} /\partial \rho = \partial f_{2D} /\partial \psi =0$. Minima of $\Delta f_{bulk}|_{\phi_0}$ satisfy the differential equation $\partial^2 \phi(z) /\partial z^2=df_{3D}/d\phi$ 
with boundary conditions $\phi(0)=\phi_0$ and $\phi(\infty)=\phi_{bulk}$. The value of $f_{2D}(\phi_0)= \min\limits_{\rho,\psi} f_{2D}$ and $\Delta f_{bulk}(\phi_0)=\min\limits_{\left\{ \phi(x) \right\}} \Delta f_{bulk}$ are plotted in \autoref{fig:LandauTheory}A, along with their sum, $f_{surf}(\phi_0)$. 
The values of $\psi$ and $\rho$ that minimize $f_{surf}(\phi_0)$ are visualized simultaneously in \autoref{fig:LandauTheory}C, each corresponding to the local minima in \autoref{fig:LandauTheory}A. 

Minima can be identified more systematically using the graphical construction in \autoref{fig:LandauTheory}B, plotting $-d f_{2D}(\phi_0) /d\phi_0$ and $\partial\Delta f_{bulk} / \partial\phi_{0}$, derivatives of the curves in 4A. 
Local minima of the surface free energy occur when these curves cross. In general, two local minima are separated by a local maximum. For two minima to have the same free energy, the area between the two curves (blue shaded regions in \autoref{fig:LandauTheory}B) must be equal.

\begin{figure}[htb!]
    \centering
    \includegraphics[width=\columnwidth]{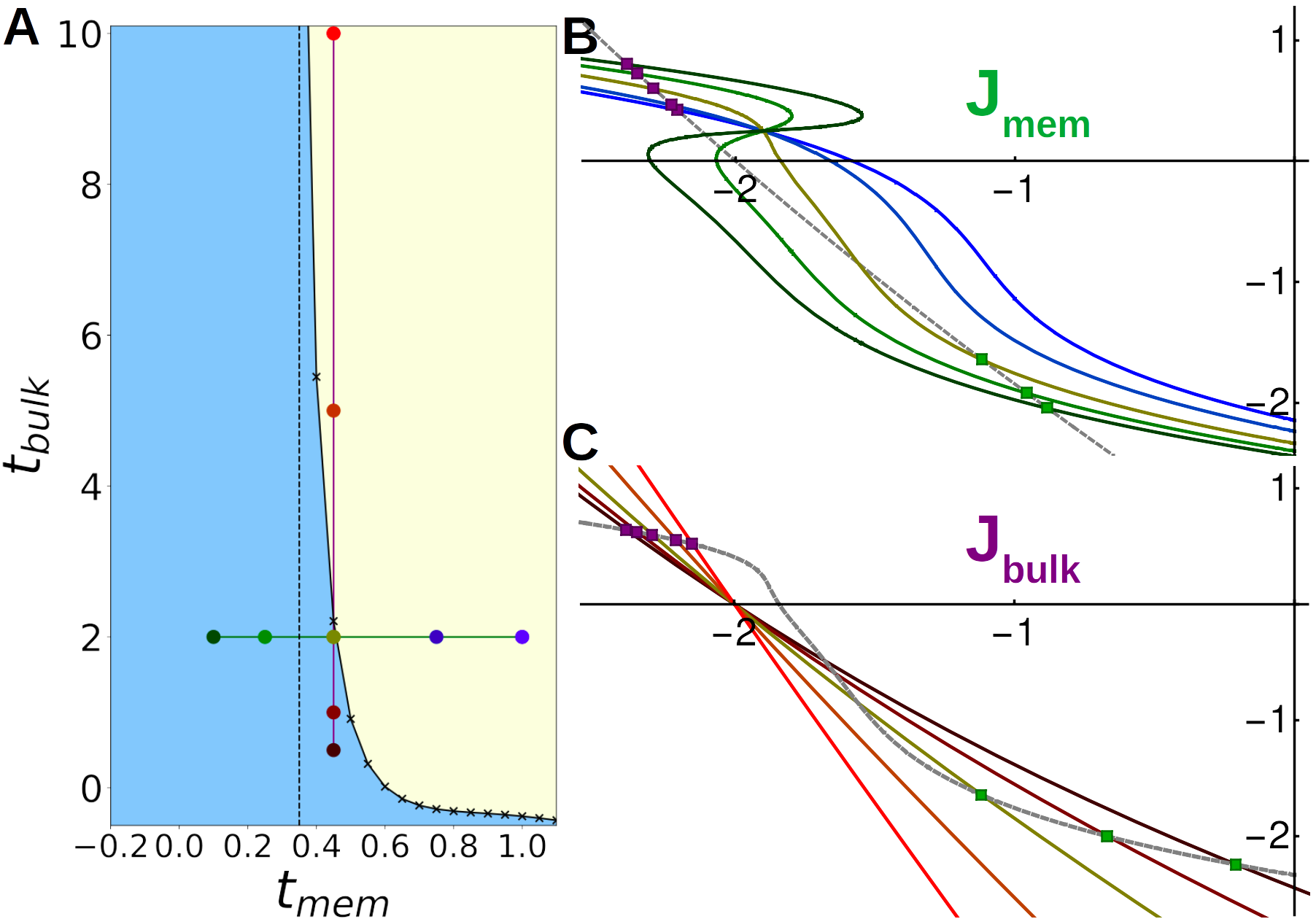}
    \caption{\textbf{Critical Point Enhancement:} \textbf{A}) Phase diagram over $J_{bulk}$ and $J_{mem}$ near the membrane critical point. The surface coexistence region (blue) extends to very weak $J_{bulk}$ near $J_{c,mem}$, marked by the dashed line. Outside of the coexistence region the surface is single-phase (yellow). \textbf{B}) Gradient construction showing how the 2D curve changes on varying membrane coupling along the the green line in A (colors from points in A). The slope of the surface curve increases as $J_{mem} \rightarrow J_{c,mem}$, diverging like the susceptibility near the membrane critical point. \textbf{C}) Gradient construction varying $J_{bulk}$ along the purple line in A. Increasing $J_{bulk}$ decreases the slope of the bulk curve, promoting surface phase coexistence.
    }
    \label{fig:Critical Point Enhancement}
\end{figure}

\noindent \textbf{Surface enhancement of bulk interactions diverges near the membrane critical point-}
We plotted the regions of surface phase coexistence as a function of bulk and membrane coupling for fixed values of $\phi_{\infty}$, $\psi$, and $\rho$, in \autoref{fig:Critical Point Enhancement}A. As with simulations we notice that the two-phase region expands significantly as $J_{mem} \rightarrow J_{c,mem}$. 

In the absence of interactions with tethers ($h_\psi=0$) the membrane of our model ($f_{mem}$) has a line of abrupt phase transitions when $t_{mem} < 0$, $\lambda_\psi=0$, terminating in a critical phase transition at $t_{mem}=0$, $\lambda_\psi=0$. For weak interactions, the location of this first order line and critical point can shift, but it's topology is unchanged - in particular, the location of the critical point shifts towards higher (positive) values of $t_{mem}$, signifying that the critical point in our simulations should occur at weaker membrane coupling. Thus the surface coexistence line should meet the membrane only transition line where $J_{bulk}=0$ as in \autoref{fig:SchematicPhaseDiagram}/\autoref{fig:Critical Point Enhancement}A, with bulk interactions supplementing membrane ones away from it. 

We can also understand the enlargement of the prewetting regime using the language of classical prewetting theories. In prewetting to a solid surface, $f_{2D}$ is typically assumed to take the simple form $f_{s} = f_0 - \mu_0\phi_0 - \frac{m_0}{2}\phi_0^2$. Here $\mu_0$ is the surface chemical potential and $m_0$ is the surface enhancement~\cite{deGennes1985} quantifying increased attractive interactions between bulk components in proximity to the surface. In most examples the surface enhancement is negative due to loss of effective interactions mediated through negative values of z. However, small positive surface enhancements are possible, for example when magnetic spins interact through contact with a surface with a larger magnetic susceptibility~\cite{Binder1988,Binder1989}. 

While our theory only explicitly includes first-order terms in $\phi_0$, higher order terms are generated by minimizing over $\psi$ and $\rho$ contributions, generating an \textit{effective surface enhancement}. In the graphical construction in \autoref{fig:LandauTheory}B, we can interpret the surface enhancement as the slope of the red $-df_{2D}/d\phi_0$ line. 
Near the critical point, components embedded in the membrane feel long range effective forces mediated by the membrane, sometimes called critical Casimir forces \cite{Reynwar2008,Machta12}. 
In surface densities these long range critical Casimir forces provide an effective surface enhancement, mediating an increased interaction between bulk components. 
The magnitude of this membrane mediated effective surface enhancement can be understood quantitatively as arising from the integral of the pairwise potential between tethers on the surface. This yields a quantity proportional to the susceptibility~\cite{Goldenfeld1992} which diverges near the critical point. This manifests as a steepening of the surface line as the membrane critical point is approached along increasing $J_{mem}$ (blue to green curves in \autoref{fig:Critical Point Enhancement}B). Below the membrane critical point we see the surface line fold back on itself, with two local minima and a local maxima at some values of $\phi_0$, implying the membrane can phase separate without bulk interactions.

While we expect our phase diagram to be topologically correct, our Landau theory fails to accurately predict the form of these phase boundaries. Mean-field theories like ours generally underestimate fluctuation effects, especially close to the critical point~\cite{Goldenfeld1992}. We expect that a more sophisticated renormalization group treatment would predict a larger criticality mediated enhancement and resulting surface coexistence region as well as a surface coexistence curve with Ising exponents rather than mean field ones. 

\begin{figure}[htb!]
    \centering
    \includegraphics[width=\columnwidth]{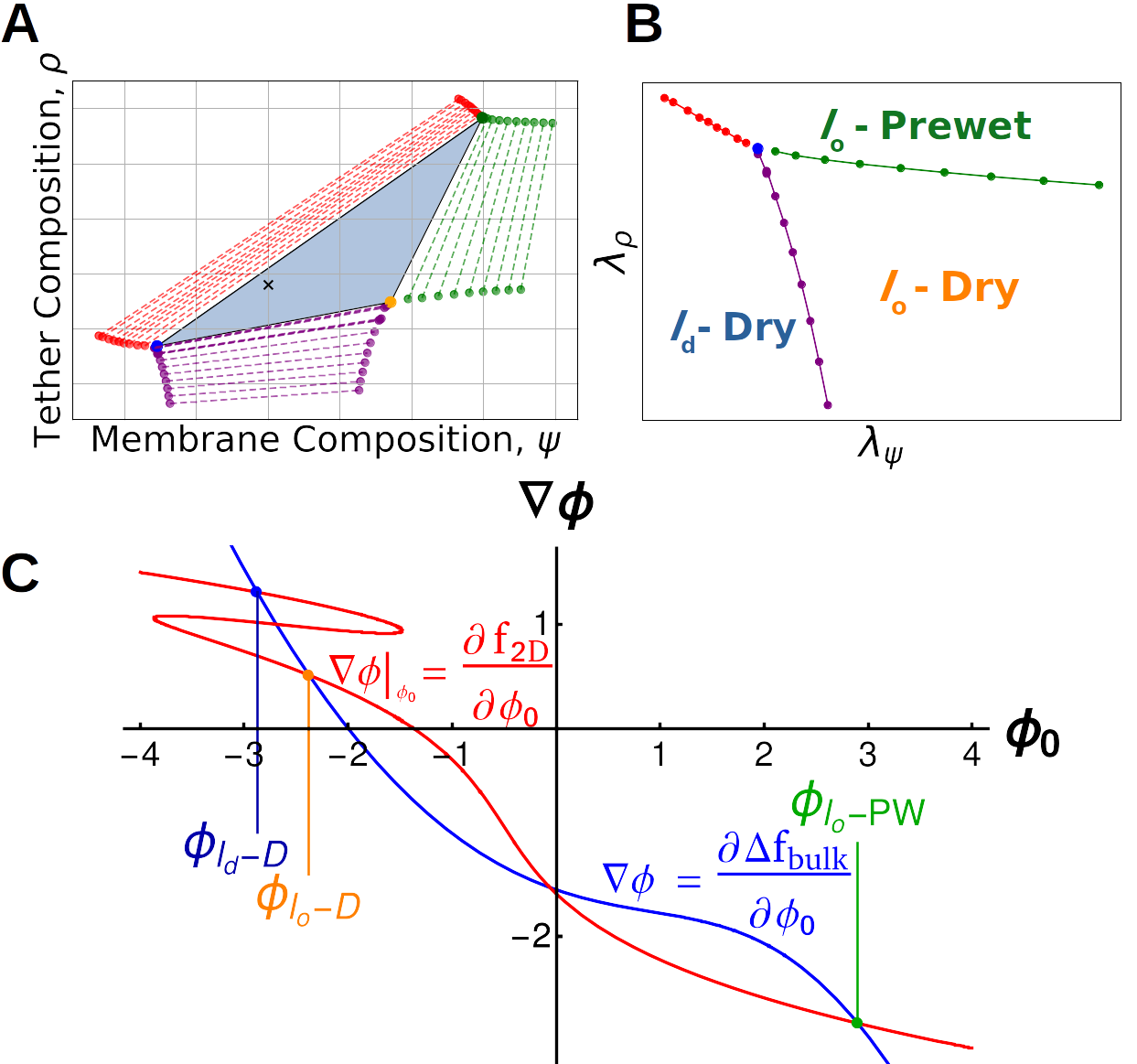}
    \caption{\textbf{Three Phase Coexistence in Landau Theory:} \textbf{A}) Phase diagram over membrane compositions ($\psi$) and tether compositions ($\rho$) calculated from Landau theory. Three phases coexist in the blue triangle, with the surface composition of each phase give by the vertices of the triangle. The positions within the triangle, (black x) sets the area fraction of phases. Three two-phase coexistence regions (red, purple, green) border the three-phase region and are plotted as tie lines. Surfaces constructed on a tie-line split into two phases with compositions given by the ends of the tie line. Single-phase regions border each two phase region. \textbf{B}) Phase diagram over the surface chemical potentials $\lambda_{\rho}$,$\lambda_{\psi}$ for the same system shown in A. The three phase triangle is represented as the blue point, and each colored line corresponds to the two phase regions in A. Outside of these lines and the point of three-phase coexistence the system is single-phase. \textbf{C}) Gradient construction within the three-phase region of A. The surface line is phase separated, folding in on itself and intersecting at the blue and yellow points. It additionally intersects with the bulk curve at high densities, green point.}
    \label{fig:Three Phase Coexistence}
\end{figure}

\noindent \textbf{Landau theory predicts coexistence of three surface phases-}
In general, each local minimum has a different value of $\psi$ and $\rho$. We expect to have two-phase coexistence when the chemical potentials $\lambda_\rho$ and $\lambda_\psi$ are such that the global minimum is doubly degenerate and three phase coexistence when the global minimum is triply degenerate. Coexistence additionally implies that the chemical potentials of each phase are identical. We minimized $\mathcal{L}$ over a range of chemical potentials searching for regions of two and three phase coexistence, shown in \autoref{fig:Three Phase Coexistence}B. We find a single point where three phases coexist and three lines of two-phase coexistence when we tune the two chemical potentials while fixing other parameters. This is permitted by Gibbs phase rule, as three phases are allowed to coexist at a single point when tuning two parameters, recapitulating the qualitative findings from our simulations.

\section*{Discussion}
We have presented a model for surface densities in which bulk components, a membrane order parameter and membrane bound tethers phase separate together in a manner reminiscent of prewetting. In our simulations the membrane is composed of a lattice Ising model, while the bulk is composed of co-acervating lattice polymers. The stability of these surface densities can be modulated by membrane interaction strength, by the density and interactions between bulk components and through the density of tethers which couple membrane and bulk. We see that when the membrane is held close to it's critical point, the regime where we see surface phase coexistence widens dramatically which we trace to membrane mediated enhancement of bulk polymer interactions. These surface densities are stable thermodynamic phases and their putative roles should be distinguished from roles the membrane may play in nucleating droplets that are already stable in the bulk but which face substantial nucleation barriers to their assembly.

While our model is not microscopically detailed, we believe it captures the coarse-grained behavior of a wide range of surface densities seen in cells. 
Building on these ideas, we propose that the unique physics of surface densities support biological function both by acting as dynamic scaffolds and by triggering cellular responses.

\noindent \textbf{Stable surface densities enable dynamic functional domains-} Prewet phases likely facilitate organization of proteins and lipids into stable, long lived complexes which perform specific functions at distinct sites. The post-synaptic density is composed of phase-separating bulk proteins~\cite{Zeng2016} adhered to a membrane domain enriched in particular ion channels, receptors and other components of the excitatory synapse. Some of these proteins, like PSD-95 are heavily palmitoylated, a modification which is dynamically regulated and confers a preference for ordered membrane lipids~\cite{Tulodziecka2016}. Palmitoylated PSD-95 likely plays a role analogous to the tethers in our model, connecting liquid ordered membrane components to cytoplasmic components of the post-synaptic density. The liquid nature of surface densities allows the post-synaptic density to dynamically exchange components with the bulk and with surrounding membrane, facilitating the mechanisms of learning which take place in neuronal synapses. 
Other structure with characteristics of surface densities play roles in neuronal mechanics. On the presynaptic side, RIM/RIM-BP condensates cluster calcium channels and machinery mediating synaptic vesicle release~\cite{Wu2019}. The inhibitory post-synaptic density displays broadly similar organization to its excitatory counterpart, but with different protein-protein interactions which instead localize inhibitory ion channels and the overlapping machinery required there~\cite{Bai2020}. 
Common across these examples are liquid structures at the membrane whose components undergo constant turnover yet whose organization and function persists over longer time scales. As in our model, the combination of membrane mediated forces and bulk interactions allows for a stable domain  highly enriched in particular components even while individual components remain mobile. 


\noindent \textbf{Surface density formation can initiate signal transduction-} Immune receptor signaling is often dependent upon membrane lipids and long-lived associations between receptors and scaffolding elements, some of which are membrane bound. Measurements in reconstituted systems mimicking T cell receptor (TCR) signaling support the idea that these scaffolds are surface densities whose formation is triggered by the phosphorylation of membrane-bound LAT~\cite{Su2016}. Some of these phosphorylations enhance interactions with soluble binding partners, equivalent to strengthening interactions with tethers within our model. Moreover, LAT is itself palmitoylated~\cite{Zhang1998}, likely conferring a $l_o$ character to LAT tethered surface densities in T cell membranes~\cite{Levental2010}. In B cell receptor signaling, it is argued that clustering receptors enhances receptor phosphorylation by stabilizing a more $l_o$-like local environment~\cite{Stone2017}. Similar to TCR, we anticipate that phosphorylation dependent interactions between membrane-bound and soluble proteins could trigger the formation of a surface density that commits to a cellular-level response. In this case, the primary function of the membrane phase transition is to enhance tether-bulk interactions via phosphorylation and not to enhance effective interactions between tethers, although this may play a role. In the Endoplasmic Reticulum (ER) the integral membrane protein IRE1 forms phase-separated clusters in response to unfolded protein stress ~\cite{Aragon2009,Belyy2020} and when lipid metabolism is disrupted~\cite{Halbleib2017}. IRE1 is thought to have an affinity for disordered lipids, but it also interacts with unfolded proteins in the ER lumen, possibly playing a role analogous to a tether in our model but in the disordered phase.

In the above examples, a signal is transduced in part by activated and sometimes cross-linked receptors seeding domains which bring downstream components into close proximity. While the specific proteins involved in these initial signaling steps are diverse, their commonality may be that in each case signal leads to increased interactions, either between membrane bound components (like increasing $J_{mem}$ in our model) between membrane and bulk components (like increasing $J_{tether}$ or $\rho$) or between particular bulk components (like changing $J_{bulk}$). In general, we propose that prewet phases serve natural roles in signaling networks owing to their unique physics. Surface transitions depend on bulk, membrane, and tether properties allowing the cell several mechanisms to regulate a single response. Moreover, prewetting can be a first-order (abrupt) transition, providing a natural mechanism to transduce a continuous signal into a discrete, switch-like cellular response.

\noindent \textbf{The interactions that drive surface densities play roles in more complex cellular structures-} Many examples in biology display some of the phenomena we investigate here but with additional subtleties and complications. Recent studies have shown that condensates can induce membrane deformations in clathrin-mediated endocytosis~\cite{Bergeron-Sandoval2018} and in synthetic systems~\cite{Yuan2020}. While we expect that surface densities can substantially deform the membrane, and that deformations may influence the interactions between membrane and bulk, we don't allow this in our model and so cannot explore it's consequences here. 
In other cases prewet phases may mediate adhesive interactions between multiple surfaces. Components of tight junctions~\cite{Beutel2019} have been shown to phase separate in the bulk, and their condensation in vivo likely includes interactions with membrane components as well as contributions from other effective forces. 
The assembly of the Golgi apparatus~\cite{Rothman2019,Rebane2020} and synaptic vesicles clustering in the presynapse~\cite{Milovanovic2018} are both thought to include proteins that phase-separate on the surface of these organelles, possibly sorting vesicles for transport. 
While our focus is on membranes, prewetting can occur on other interesting biological surfaces. Phase separation has been proposed to play prominent roles in transcriptional regulation~\cite{Hnisz2017,Bialek2019}, where DNA has been proposed to act as a one-dimensional `surface' for prewetting~\cite{Morin2020}. 

\noindent \textbf{Surface densities can be driven by membrane or bulk interactions alone or through a combination-} The lipid composition of plasma membranes appears to be tuned close to a thermodynamic critical point ~\cite{Veatch08,Gray2013,Gray2015}, which we have argued has important consequences for surface phases. Near the membrane critical point, the bulk coupling needed to see surface coexistence rapidly decreases (\autoref{fig:Critical Point Enhancement}). 
We expect that surface densities could be stabilized entirely through membrane criticality mediated interactions, solely through prewetting interactions between bulk components, or through a mixture of these two forces. The two extremes have been explored in synthetic systems. Synthetic membranes can phase separate into coexisting two-dimensional liquid phases in the absence of any proteins. More recently, two-dimensional coexisting phases have been observed on single-component membranes~\cite{Last2020,Love2020}, driven by interactions between bulk proteins some of which adhere. Similar experiments in multi-component membranes~\cite{Lee2019,Chung2020} highlight the bulk's ability to mediate interactions between membrane lipids. 
Because these interactions are stable outside of the regime of bulk coexistence, they are most likely prewet. In cells, proximity of the membrane to its critical point likely allows for weak and diverse interactions between sparse proteins leading to surface phases far outside of their coexistence regime and even far outside of the regime in which they would prewet a single component membrane.

\noindent \textbf{Prewetting appears to be more common than wetting in cellular phase separation-} The phase-separated $\beta$-catenin destruction complex~\cite{Schaefer2019}, integral to Wnt signaling, is recruited to the plasma membrane on induction of the Wnt pathway, remaining a dynamic assembly on the membrane~\cite{schwarz-romond_wnt_2005}. This is likely analogous to a transition between a dry and wet phases in our model. However, nearly all recently described cytoplasmic condensates are observed away from membranes~\cite{Alberti2017,Shin2017}, even though our model suggests that only weak, tether-driven interactions are required for membrane wetting. By contrast, a large number of cellular structures appear to be prewet - forming thin films on specific membrane domains outside of bulk coexistence.
This may suggest that attractive interactions between droplets and cytoskeletal elements outcompete interactions with membrane components, or that these interactions are limited by material properties of the cortex~\cite{Ronceray2021}.

The prediction of prewetting~\cite{Cahn1977} significantly preceded its first experimental realizations~\cite{Rutledge1992,Kellay1993}. Prewet phases outside of biology typically require fine-tuning and subtle experimental considerations to observe. By contrast, in biological contexts, surface densities appear to be common, owing to the presence of a complex membrane with a propensity to phase separate interacting with a dense polymer solution. Our conception of surface densities includes membrane dominated phases, close to the usual concept of a lipid raft, bulk driven phases that closely match the classical concept of prewetting, as well as phases which make use of a combination of these interactions. We hope that future work will clarify the roles these surface densities play in diverse cellular functions.


\matmethods{Simulation code, Landau Theory calculations, and supplemental text and videos can be found on GitHub at \href{https://github.com/SimludDalhec/critical-membrane-prewetting}{critical membrane prewetting}

\noindent{\textbf{Monte-Carlo Simulations}}
Monte-Carlo simulations were implemented on a 3-Dimensional lattice ($D\text{x}L\text{x}L$) populated with polymers, tethers, and a membrane simulated by an Ising model. The lattice is periodic in the two $L$ dimensions and has free boundary conditions at $D=0,L$, with the Ising model located at the $D = 0$ boundary. Our model is described by a simple Hamiltonian:

\begin{align}
    \mathcal{H}_{bulk} &= J_{bulk}\sum_{i}\sigma^{blue}_{i}\sigma^{red}_{i} + J_{nn}\sum_{i,j \in nn}{\sigma_{i}\sigma_{j}} - \mu_{bulk}N_{bulk}\nonumber\\
    \mathcal{H}_{ising} &= J_{ising}\sum_{i,j \in nn}{s_{i}s_{j}}\nonumber\\
    \mathcal{H}_{tether} &= J_{tether}\sum_{i \in tethers}\sigma^{bulk}_{i}\sigma^{tether}_{i}
\end{align}

Where $J_{bulk}$ is the interaction strength between polymers of different types (`red' and `blue'), $J_{nn}$ is a nearest neighbor energy, and $\mu_{bulk}$ is the chemical potential of the 3D system. The spins within the Ising model interact with coupling $J_{ising}$ and components of the bulk interact with  tethers through $J_{tether}$
 
\noindent\textbf{Bulk Polymers:} Cytoplasmic proteins are simulated as a mixture of lattice polymers. Bulk polymers occupy the vertices of a 3D cubic lattice. Snake-like moves where the tail of the polymer is moved to a free space adjacent to the head (and vice-versa) allow polymers to explore the lattice. Here we simulate just two bulk polymer species and a single tether species. Polymers of the same type cannot inhabit the same lattice position while polymers of opposite type interact through $J_{bulk}$ when occupying the same lattice site. All bulk polymers interact equally with tethers. Additionally, all polymers and tethers have a small, favorable nearest-neighbor interactions $J_{nn} = 0.1k_{b}T$. This nearest-neighbor energy is required to give the droplets tension, without which they do not condense
~\cite{Freeman_Rosenzweig2017,Xu2020}.

\noindent\textbf{Tethers:} Tethers move in two dimensions across the surface of the Ising model. Proposed moves translate a tether one lattice space in a random direction. Proposals that move the tether off of an up spin or result in two tethers occupying the same lattice site are immediately rejected. 

\noindent\textbf{Membrane:} The membrane is simulated as a conserved order parameter Ising model, implemented on a 2D cubic lattice with periodic boundary conditions. To conserve the total magnetization, or lipid composition, we use a non-local Kawasaki moves where Ising spins are exchanged rather than flipped. We fix up-spins at every tether-occupied site during each sweep.

\noindent\textbf{Simulation Scheme:} Each simulation consists of sweeps through polymers, tethers, and membrane spins. We proposed moves through a randomized sequence of polymers and tethers in the system, followed by a sweep through all Ising spins, and proposal of particle exchanges. We equilibrate simulations by raising bulk coupling and tether coupling in increments of $0.05 - 0.10 k_{b}T$ with $1\times10^5$ - $5\times10^6$ Monte-Carlo sweeps per temperature step. Simulations were sometimes extended from the previous endpoints, for up to $5\times10^6$ Monte-Carlo sweeps at a single set of parameters, to ensure equilibration.

Single polymer, tether, and Ising moves are accepted with the Metropolis probability $e^{-(H_f - H_i)/k_bT}$ where $H_f$,$H_i$ are the energies of the final and initial system configurations. To accelerate equilibration we propose cluster moves where a connected set of polymers translate one lattice spacing. Cluster moves are proposed with probability $(1 / N_{poly})$ and are only accepted if the move does not form or break any bonds, satisfying detailed balance.

In simulations at fixed $\mu_{bulk}$, polymers are exchanged between the system and a non-interacting reservoir. The amount of polymers exchanged per Monte Carlo step is sampled from a Poisson distribution where $\lambda = \frac{N_{sys} + N_{res}}{N_{init} + N_{res}}$, ensuring that chemical potential remains constant as particles are added to the system. Exchanges are accepted with probability $e^{-\Delta H_{nn} - \mu_{bulk}}$, where $\Delta H_{nn}$ is the change in nearest neighbor energy. Exchanges that remove or add bonds to the system are immediately rejected. Swapping a particle from the reservoir to the system simply copies the reservoir particle into the system while moving a particle from the system to the reservoir removes the particle from the system but does not place an additional particle in the reservoir. This scheme of `virtual' exchanges is done so the reservoir is effectively infinite while we only simulate a finite amount of particles.

\noindent\textbf{Extracting Surface Composition from Simulations}: To obtain the membrane and tether compositions of simulations that appeared to have 3 coexisting phases, we analyzed histograms of membrane and tether composition. First we averaged the membrane spins and tether positions over 50000 MCS. From this time-average, we scanned the surface with a 5x5 grid, computing the average membrane and tether compositions within. These values are collected over the last half of simulation run, 2,500,000 MCS, and used to construct a two-dimensional histogram of tether and membrane compositions. We defined the surface composition of coexisting phases as the peaks of this histogram. Because there were multiple peaks likely corresponding to a single phase, we required that the difference in tether density between peaks was greater than 0.05.\\
\noindent\textbf{{Simulation Parameters used in figures}}
\begin{center}
\resizebox{\columnwidth}{!} {%
\begin{tabular}{ |c|c|c|c|c|c|c| } 
 \hline
 \textbf{Figure} & \textbf{Tether Density} & \textbf{Membrane Order} & $J_{bulk},k_{B}T$ & $J_{mem}, T_{c}$ & $J_{tether}$ & \textbf{$\mu_{bulk}$/$\phi_{bulk}$} \\\hline 
 1D, Dry & 0.0 & 0.5 & 1.0 & 1.0 & 0.0 & 0.06\\ 
 1D, Wet & 0.03125 & 0.5 & 0.9 & 1.1 & 0.5 & 0.10 \\ 
 1D, Surface Coexistence & 0.0468 & 0.5 & 0.75 & 1.0 & 1.0 & -5.9\\
 1D, Single Phase & 0.0468 & 0.5 & 0.675 &1.0 & 1.0 & -5.9\\
 2A& 0.0468 & 0.5 & 0.85 & 1.05 &  1.0 &-5.9 \\
 2C, Center & 0.0468 & 0.5 & 0.775 & 2.0 & 1.0 & -5.9\\
 2C, Upper & 0.0468 & 0.5 & 0.825 & 2.0 & 1.0 & -5.9\\
 2C, Right & 0.0468 & 0.5 & 0.775 & 1.0  & 1.0 &-5.9\\
 3A  & 0.0625 & 0.6 & 0.5 & 0.9 & 1.0 & -4.5\\
 \hline
\end{tabular}
}
\end{center}
\noindent\textbf{{Mean-Field Theory}}: To minimize the free-energy of our system we sought to express the contributions from bulk terms in terms of surface and bulk densities $\phi_0$ and $\phi_{\infty}$, as these alone determine the density profile. Following previous work~\cite{Cahn1977,deGennes1985}, we identify spatial gradients $\nabla\phi$ with distance from $\phi_{bulk}$
\begin{equation*}
    \nabla\phi =  \pm\sqrt{2(f_{3D}(\phi_{0}) - f_{bulk})}
\end{equation*}
Where this follows from the functional derivative $\frac{\delta L_{3D}}{\delta\phi_{z}}$. We use this identity to express the contributions from spatial variations of $\Delta\phi(z)$ in terms of $\phi_0$ and $\phi_{bulk}$

\begin{align}
        \Delta f_{bulk} &= \int_0^{\infty}{dz \frac{1}{2}(\nabla\phi)^2 + f_{3D}(\phi) - f_{bulk}}\nonumber\\
         &= \int_0^{\infty}{dz \left(\frac{d\phi}{dz}\right) \left(\frac{dz}{d\phi}\right)  \frac{1}{2}(\nabla\phi)^2 + f_{3D}(\phi) - f_{bulk}}\nonumber\\
         &= \int_{\phi(0)}^{\phi(\infty)}{d\phi (\nabla\phi)^{-1}\frac{1}{2}(\nabla\phi)^2 + f_{3D}(\phi) - f_{bulk}}\nonumber\\
         &= \int_{\phi_0}^{\phi_{bulk}}{d\phi (\nabla\phi)^{-1}\frac{1}{2}(\nabla\phi)^2 + \underbrace{f_{3D}(\phi) - f_{bulk}}_{\frac{1}{2}\nabla\phi}}\nonumber\\
         &= \int_{\phi_0}^{\phi_{bulk}}{d\phi (\nabla\phi)^{-1}(\nabla\phi)^2}\nonumber\\
    	\Delta f_{bulk}(\phi_0,\phi_{\infty}) &= \int_{\phi_0}^{\phi_{bulk}}{d\phi \sqrt{2(f_{3D}(\phi_{0}) - f_{bulk}))}}
\end{align}
The total free energy of the bulk and surface terms, $f_{surf}$, can now be written as:
\begin{equation}
    f_{surf} =f_{2D}(\rho,\psi,\phi_0) + \int_{\phi_0}^{\phi_{bulk}}{d\phi \sqrt{2(f_{3D}(\phi_{0}) - f_{bulk}))}}
\end{equation}
Which we minimize numerically over values of $\phi_0$, $\psi$, $\rho$ to obtain results throughout the  text. $f_{surf}$ can be minimized independently over $\phi_0$,$\psi$, or $\rho$ values to obtain the surface free energy as a function of the remaining terms, as plotted in \autoref{fig:LandauTheory}A,C. 

\noindent\textbf{Numerical Phase Diagrams}: We minimized $f_{surf}$ numerically with Mathematica. We calculated solutions at over a range $\lambda_{\rho}$,$\lambda_{\psi}$ values to find coexistence regions. When there were multiple solutions with near-identical energies, within $< 0.1 k_{b}T$, we declared them coexisting phases. Values of $\psi$,$\rho$ that minimize the free energy at these points terminate tie lines in a fixed composition system. This procedure is visualized in \autoref{fig:Three Phase Coexistence}A,B where the $\psi$, $\rho$ values in A correspond to $\lambda_{\rho}$,$\lambda_{\psi}$ values in B. Multiple phase diagrams in the space of $J_{bulk}$,$J_{mem}$ were constructed through combining the tie lines and three phase regions of of phase diagrams calculated at values of $t_{mem}$ and $t_{bulk}$. At a specific $\psi$, $\rho$ values we determined whether the system was in a one, two, or three phase region of the phase diagram. 

\noindent\textbf{Landau theory parameters used in figures}\\
\begin{center}
\resizebox{\columnwidth}{!} {
\begin{tabular}{ |c|c|c|c|c|c|c|c|c|c|c| } 
 \hline
 \textbf{Figure} & $t_{mem}$ & $t_{bulk}$ & $\phi_{\infty}$ & $\rho_{\star}$ & $\lambda_{\rho}$ & $\lambda_{\psi}$ & $\rho$ & $\psi$ & $h_{\phi}$ & $h_{\psi}$ \\\hline 
 Figure 4 & 1.1 & -0.1 & -2.0 & 1 & -0.1 & 0.5 & N/A & N/A & 1 & 1 \\ 
 Figure 5A & N/A & N/A & -2.0 & 1 & N/A & N/A & -0.5 & 0 & 1 & 1 \\
 Figure 5B & 1, 0.75, 0.45, 0.25, 0.1 & 2 & -2.0 & 1 & -0.05 & 0.25 & N/A & N/A & 1 & 1 \\
 Figure 5C & 0.45 & 10, 5, 2, 1, 0.5 & -2.0 & 1 & -0.05 & 0.25 & N/A & N/A & 1 & 1 \\
 Figure 6A,B & -0.2 & -0.4 & -2.0 & 1 & N/A & N/A & N/A & N/A & 1 & 1 \\
 Figure 6C & -0.2 & -0.4 & -2.0 & 1 & -2 & 1 & N/A & N/A & 1 & 1 \\
 \hline
\end{tabular}}
\end{center}
}
\showmatmethods{} 

\acknow{We thank Isabella Graf and Jon Machta for helpful comments on a draft, and Ilya Levental and Hong-Yin Wang for useful discussions. This work was supported by NSF BMAT 1808551 (MR, SLV and BBM), NIH R35 GM138341 (BBM) and NSF 1522467 (MR).} 

\showacknow{} 

\bibliography{refs.bib}

\begin{thebibliography}{10}

\bibitem{Alberti2017}
S Alberti, Phase separation in biology.
\newblock {\em\protect\JournalTitle{Current Biology}} \textbf{27}, R1097--R1102
  (2017).

\bibitem{Sengupta07}
P Sengupta, B Baird, D Holowka, Lipid rafts, fluid/fluid phase separation, and
  their relevance to plasma membrane structure and function.
\newblock {\em\protect\JournalTitle{Seminars in Cell \& Developmental Biology}}
  \textbf{18}, 583--590 (2007).

\bibitem{Brangwynne09}
CP Brangwynne, et~al., Germline {P} {Granules} {Are} {Liquid} {Droplets} {That}
  {Localize} by {Controlled} {Dissolution}/{Condensation}.
\newblock {\em\protect\JournalTitle{Science}} \textbf{324}, 1729--1732 (2009).

\bibitem{Li2012}
P Li, et~al., Phase transitions in the assembly of multivalent signalling
  proteins.
\newblock {\em\protect\JournalTitle{Nature}} \textbf{483}, 336--340 (2012).

\bibitem{Priftis2012}
D Priftis, M Tirrell, Phase behaviour and complex coacervation of aqueous
  polypeptide solutions.
\newblock {\em\protect\JournalTitle{Soft Matter}} \textbf{8}, 9396--9405
  (2012).

\bibitem{Honerkamp-Smith09}
AR Honerkamp-Smith, SL Veatch, SL Keller, An introduction to critical points
  for biophysicists; observations of compositional heterogeneity in lipid
  membranes.
\newblock {\em\protect\JournalTitle{Biochimica et Biophysica Acta -
  Biomembranes}} \textbf{1788}, 53--63 (2009).

\bibitem{Veatch08}
SL Veatch, et~al., Critical fluctuations in plasma membrane vesicles.
\newblock {\em\protect\JournalTitle{ACS Chem Bio}} \textbf{3}, 287--293 (2008).

\bibitem{Banjade2014}
S Banjade, MK Rosen, Phase transitions of multivalent proteins can promote
  clustering of membrane receptors.
\newblock {\em\protect\JournalTitle{eLife}} \textbf{3}, e04123 (2014).

\bibitem{Su2016}
X Su, et~al., Phase separation of signaling molecules promotes {T} cell
  receptor signal transduction.
\newblock {\em\protect\JournalTitle{Science}} \textbf{352}, 595--599 (2016).

\bibitem{Zeng2016}
M Zeng, et~al., Phase {Transition} in {Postsynaptic} {Densities} {Underlies}
  {Formation} of {Synaptic} {Complexes} and {Synaptic} {Plasticity}.
\newblock {\em\protect\JournalTitle{Cell}} \textbf{166}, 1163--1175 (2016).

\bibitem{Beutel2019}
O Beutel, R Maraspini, K Pombo-García, Cl Martin-Lemaitre, A Honigmann, Phase
  {Separation} of {Zonula} {Occludens} {Proteins} {Drives} {Formation} of
  {Tight} {Jun\ ctions}.
\newblock {\em\protect\JournalTitle{Cell}} \textbf{179}, 923--936.e11 (2019).

\bibitem{Wu2019}
X Wu, et~al., {RIM} and {RIM}-{BP} {Form} {Presynaptic} {Active}-{Zone}-like
  {Condensates} via {Phase} {Separation}.
\newblock {\em\protect\JournalTitle{Molecular Cell}} \textbf{73}, 971--984.e5
  (2019).

\bibitem{Zeng2018}
M Zeng, et~al., Reconstituted {Postsynaptic} {Density} as a {Molecular}
  {Platform} for {Understanding} {Synapse} {Formation} and {Plasticity}.
\newblock {\em\protect\JournalTitle{Cell}} \textbf{174}, 1172--1187 (2018).

\bibitem{Case2019}
LB Case, X Zhang, JA Ditlev, MK Rosen, Stoichiometry controls activity of
  phase-separated clusters of actin signaling proteins.
\newblock {\em\protect\JournalTitle{Science}} \textbf{363}, 1093--1097 (2019).

\bibitem{deGennes1985}
PG de~Gennes, Wetting: statics and dynamics.
\newblock {\em\protect\JournalTitle{Reviews of Modern Physics}} \textbf{57},
  827--863 (1985).

\bibitem{Bonn2009}
D Bonn, J Eggers, J Indekeu, J Meunier, E Rolley, Wetting and spreading.
\newblock {\em\protect\JournalTitle{Reviews of Modern Physics}} \textbf{81},
  739--805 (2009).

\bibitem{Cahn1977}
JW Cahn, Critical point wetting.
\newblock {\em\protect\JournalTitle{The Journal of Chemical Physics}}
  \textbf{66}, 3667--3672 (1977).

\bibitem{Nakanishi1982}
H Nakanishi, ME Fisher, Multicriticality of {Wetting}, {Prewetting}, and
  {Surface} {Transitions}.
\newblock {\em\protect\JournalTitle{Physical Review Letters}} \textbf{49},
  1565--1568 (1982).

\bibitem{Brangwynne2015}
CP Brangwynne, P Tompa, RV Pappu, Polymer physics of intracellular phase
  transitions.
\newblock {\em\protect\JournalTitle{Nature Physics}} \textbf{11}, 899--904
  (2015).

\bibitem{Honerkamp-Smith08}
AR Honerkamp-Smith, et~al., Line {Tensions}, {Correlation} {Lengths}, and
  {Critical} {Exponents} in {Lip\ id} {Membranes} {Near} {Critical} {Points}.
\newblock {\em\protect\JournalTitle{Biophysical Journal}} \textbf{95}, 236--246
  (2008).

\bibitem{Machta11}
BB Machta, S Papanikolaou, JP Sethna, SL Veatch, Minimal model of plasma
  membrane heterogeneity requires coupling cortical actin to criticality.
\newblock {\em\protect\JournalTitle{Biophys J}} \textbf{100}, 1668--1677
  (2011).

\bibitem{Freeman_Rosenzweig2017}
ES Freeman~Rosenzweig, et~al., The {Eukaryotic} {CO2}-{Concentrating}
  {Organelle} {Is} {Liquid}-like and {Exhibits} {Dynamic} {Reorganization}.
\newblock {\em\protect\JournalTitle{Cell}} \textbf{171}, 148--162.e19 (2017).

\bibitem{Xu2020}
B Xu, et~al., Rigidity enhances a magic-number effect in polymer phase
  separation.
\newblock {\em\protect\JournalTitle{Nature Communications}} \textbf{11}, 1561
  (2020).

\bibitem{priftis2012-1}
D Priftis, M Tirrell, Phase behaviour and complex coacervation of aqueous
  polypeptide solutions.
\newblock {\em\protect\JournalTitle{Soft Matter}} \textbf{8}, 9396--9405
  (2012).

\bibitem{priftis2012-2}
D Priftis, N Laugel, M Tirrell, Thermodynamic {Characterization} of
  {Polypeptide} {Complex} {Coacervation}.
\newblock {\em\protect\JournalTitle{Langmuir}} \textbf{28}, 15947--15957
  (2012).

\bibitem{Snead2019}
WT Snead, AS Gladfelter, The {Control} {Centers} of {Biomolecular} {Phase}
  {Separation}: {How} {Membrane} {Surfaces}, {PTMs}, and {Active} {Processes}
  {Regulate} {Condensation}.
\newblock {\em\protect\JournalTitle{Molecular Cell}} \textbf{76}, 295--305
  (2019).

\bibitem{Shin2017}
Y Shin, CP Brangwynne, Liquid phase condensation in cell physiology and
  disease.
\newblock {\em\protect\JournalTitle{Science}} \textbf{357}, eaaf4382 (2017).

\bibitem{Jiang2018}
H Jiang, et~al., Protein {Lipidation}: {Occurrence}, {Mechanisms}, {Biological}
  {Functions}, and {Enabling}\ {Technologies}.
\newblock {\em\protect\JournalTitle{Chemical Reviews}} \textbf{118}, 919--988
  (2018).

\bibitem{Goldenfeld1992}
N Goldenfeld, {\em Lectures on phase transitions and the renormalization
  group}, Frontiers in physics.
\newblock (Addison-Wesley, Advanced Book Program, Reading, Mass) No.{} v. 85,
  (1992).

\bibitem{Binder1988}
K Binder, DP Landau, Wetting and layering in the nearest-neighbor simple-cubic
  {Ising} lattice: {A} {Monte} {Carlo} in\ vestigation.
\newblock {\em\protect\JournalTitle{Physical Review B}} \textbf{37}, 1745--1765
  (1988).

\bibitem{Binder1989}
K Binder, DP Landau, S Wansleben, Wetting transitions near the bulk critical
  point: {Monte} {Carlo} simulations for the {Ising} model.
\newblock {\em\protect\JournalTitle{Physical Review B}} \textbf{40}, 6971--6979
  (1989).

\bibitem{Reynwar2008}
BJ Reynwar, M Deserno, Membrane composition-mediated protein-protein
  interactions.
\newblock {\em\protect\JournalTitle{Biointerphases}} \textbf{3}, FA117--FA124
  (2008).

\bibitem{Machta12}
BB Machta, SL Veatch, JP Sethna, Critical casimir forces in cellular membranes.
\newblock {\em\protect\JournalTitle{Phys Rev Lett}} \textbf{109}, 138101
  (2012).

\bibitem{Tulodziecka2016}
K Tulodziecka, et~al., Remodeling of the postsynaptic plasma membrane during
  neural development.
\newblock {\em\protect\JournalTitle{Molecular Biology of the Cell}}
  \textbf{27}, 3480--3489 (2016).

\bibitem{Bai2020}
G Bai, Y Wang, M Zhang, Gephyrin-mediated formation of inhibitory postsynaptic
  density sheet via phase separation.
\newblock {\em\protect\JournalTitle{Cell Research}} (2020).

\bibitem{Zhang1998}
W Zhang, RP Trible, LE Samelson, {LAT} {Palmitoylation}.
\newblock {\em\protect\JournalTitle{Immunity}} \textbf{9}, 239--246 (1998).

\bibitem{Levental2010}
I Levental, D Lingwood, M Grzybek, U Coskun, K Simons, Palmitoylation regulates
  raft affinity for the majority of integral raft proteins.
\newblock {\em\protect\JournalTitle{Proceedings of the National Academy of
  Sciences}} \textbf{107}, 22050--22054 (2010).

\bibitem{Stone2017}
MB Stone, SA Shelby, MF Núñez, K Wisser, SL Veatch, Protein sorting by lipid
  phase-like domains supports emergent signaling function in b lymphocyte
  plasma membranes.
\newblock {\em\protect\JournalTitle{eLife}} \textbf{6}, e19891 (2017).

\bibitem{Aragon2009}
T Aragón, et~al., Messenger {RNA} targeting to endoplasmic reticulum stress
  signalling sites.
\newblock {\em\protect\JournalTitle{Nature}} \textbf{457}, 736--740 (2009).

\bibitem{Belyy2020}
V Belyy, NH Tran, P Walter, Quantitative microscopy reveals dynamics and fate
  of clustered {IRE1$\alpha$}.
\newblock {\em\protect\JournalTitle{Proceedings of the National Academy of
  Sciences}} \textbf{117}, 1533--1542 (2020).

\bibitem{Halbleib2017}
K Halbleib, et~al., Activation of the {Unfolded} {Protein} {Response} by
  {Lipid} {Bilayer} {Stress}.
\newblock {\em\protect\JournalTitle{Molecular Cell}} \textbf{67}, 673--684.e8
  (2017).

\bibitem{Bergeron-Sandoval2018}
LP Bergeron-Sandoval, et~al., Endocytosis caused by liquid-liquid phase
  separation of proteins.
\newblock {\em\protect\JournalTitle{BioRxiv}} (2017).

\bibitem{Yuan2020}
F Yuan, et~al., Membrane bending by protein phase separation.
\newblock {\em\protect\JournalTitle{BioRxiv}}, 22 (2020).

\bibitem{Rothman2019}
JE Rothman, Jim's {View}: {Is} the {Golgi} stack a phase-separated liquid
  crystal?
\newblock {\em\protect\JournalTitle{FEBS Letters}} \textbf{593}, 2701--2705
  (2019).

\bibitem{Rebane2020}
AA Rebane, et~al., Liquid–liquid phase separation of the {Golgi} matrix
  protein {GM130}.
\newblock {\em\protect\JournalTitle{FEBS Letters}} \textbf{594}, 1132--1144
  (2020).

\bibitem{Milovanovic2018}
D Milovanovic, Y Wu, X Bian, P De~Camilli, A liquid phase of synapsin and lipid
  vesicles.
\newblock {\em\protect\JournalTitle{Science}} \textbf{361}, 604--607 (2018).

\bibitem{Hnisz2017}
D Hnisz, K Shrinivas, RA Young, AK Chakraborty, PiA Sharp, A {Phase}
  {Separation} {Model} for {Transcriptional} {Control}.
\newblock {\em\protect\JournalTitle{Cell}} \textbf{169}, 13--23 (2017).

\bibitem{Bialek2019}
W Bialek, T Gregor, G Tkačik, Action at a distance in transcriptional
  regulation.
\newblock {\em\protect\JournalTitle{arXiv:1912.08579 [cond-mat,
  physics:physics, q-bio]}} (2019) arXiv: 1912.08579.

\bibitem{Morin2020}
JA Morin, et~al., Surface condensation of a pioneer transcription factor on
  {DNA}.
\newblock {\em\protect\JournalTitle{BioRxiv}} (2020).

\bibitem{Gray2013}
E Gray, J Karslake, BB Machta, SL Veatch, Liquid {General} {Anesthetics}
  {Lower} {Critical} {Temperatures} in {Plasma} {Membrane} {Vesicles}.
\newblock {\em\protect\JournalTitle{Biophysical Journal}} \textbf{105},
  2751--2759 (2013).

\bibitem{Gray2015}
EM Gray, G Díaz-Vázquez, SL Veatch, Growth {Conditions} and {Cell} {Cycle}
  {Phase} {Modulate} {Phase} {Transition} {Temperatures} in {RBL}-{2H3}
  {Derived} {Plasma} {Membrane} {Ves\ icles}.
\newblock {\em\protect\JournalTitle{PLOS ONE}} \textbf{10}, e0137741 (2015).

\bibitem{Last2020}
MGF Last, S Deshpande, C Dekker, {pH}-{Controlled} {Coacervate}–{Membrane}
  {Interactions} within {Liposomes}.
\newblock {\em\protect\JournalTitle{ACS Nano}} \textbf{14}, 4487--4498 (2020).

\bibitem{Love2020}
C Love, et~al., Reversible {pH}-{Responsive} {Coacervate} {Formation} in
  {Lipid} {Vesicles} {Activates} {Dormant} {Enzymatic} {Reactions}.
\newblock {\em\protect\JournalTitle{Angewandte Chemie International Edition}}
  \textbf{59}, 5950--5957 (2020).

\bibitem{Lee2019}
IH Lee, MY Imanaka, EH Modahl, AP Torres-Ocampo, Lipid {Raft} {Phase}
  {Modulation} by {Membrane}-{Anchored} {Proteins} with {Inherent} {Phase}
  {Sepa\ ration} {Properties}.
\newblock {\em\protect\JournalTitle{ACS Omega}} \textbf{4}, 6551--6559 (2019).

\bibitem{Chung2020}
JK Chung, et~al., Coupled membrane lipid miscibility and phosphotyrosine-driven
  protein conden\ sation phase transitions.
\newblock {\em\protect\JournalTitle{Biophysical Journal}} \textbf{119},
  S0006349520307281 (2020).

\bibitem{Schaefer2019}
KN Schaefer, M Peifer, Wnt/{Beta}-{Catenin} {Signaling} {Regulation} and a
  {Role} for {Biomolecular} {Condensates}.
\newblock {\em\protect\JournalTitle{Developmental Cell}} \textbf{48}, 429--444
  (2019).

\bibitem{schwarz-romond_wnt_2005}
T Schwarz-Romond, C Merrifield, BJ Nichols, M Bienz, The {Wnt} signalling
  effector {Dishevelled} forms dynamic protein assemblies rather than stable
  associations with cytoplasmic vesicles.
\newblock {\em\protect\JournalTitle{Journal of Cell Science}} \textbf{118},
  5269--5277 (2005).

\bibitem{Ronceray2021}
P Ronceray, S Mao, A Košmrlj, MP Haataja, Liquid demixing in elastic networks:
  cavitation, permeation, or size selecti\ on?
\newblock {\em\protect\JournalTitle{arXiv:2102.02787 [cond-mat,
  physics:physics]}} (2021) arXiv: 2102.02787.

\bibitem{Rutledge1992}
JE Rutledge, P Taborek, Prewetting phase diagram of {He} 4 on cesium.
\newblock {\em\protect\JournalTitle{Physical Review Letters}} \textbf{69},
  937--940 (1992).

\bibitem{Kellay1993}
H Kellay, D Bonn, J Meunier, Prewetting in a binary liquid mixture.
\newblock {\em\protect\JournalTitle{Physical Review Letters}} \textbf{71},
  2607--2610 (1993).

\end{thebibliography}

\end{document}